\documentclass[
	fontsize=12pt,
	oneside,
	DIV=classic,
	paper=a4,
	pagesize=auto,
	titlepage=yes,
	bibliography		= totocnumbered,
	listof					= totocnumbered,
	parskip					= half,
	abstract				= yes
	]{scrartcl}

	\usepackage[utf8]{inputenc}
	\usepackage[T1]{fontenc}
	\usepackage[english]{babel}
	
	\usepackage{times}
	\addtokomafont{descriptionlabel}{\rmfamily}
	\addtokomafont{pagehead}{\scshape}

		\usepackage[activate={true,nocompatibility},final,tracking=false,kerning=true,spacing=true,factor=1100,stretch=10,shrink=10]{microtype}
	
	\usepackage{enumerate}
				
	\usepackage{csquotes}
	\usepackage[backend=bibtex8,style=numeric,autolang=other]{biblatex}
	
	\bibliography{literature}

	\usepackage[headsepline]{scrlayer-scrpage}
	\ohead{}
	\ihead{Implementing a financial derivative as smart contract}
	\usepackage[intlimits]{amsmath}
	\usepackage{algorithmic,algorithm,listings}
	
	\usepackage{graphicx,xcolor,subfig}
	
	\definecolor{DarkBlue}{rgb}{0.0, 0.0, 0.5}
	\definecolor{DarkRed}{rgb}{0.5, 0.0, 0.0}
	\definecolor{DarkGreen}{rgb}{0.0, 0.5, 0.0}
	\definecolor{DarkYellow}{rgb}{0.5, 0.5, 0.0}
	\definecolor{Brown}{cmyk}{0.00,0.80,1.00,0.60}
	\definecolor{DarkGreen}{cmyk}{0.64,0.00,0.95,0.60}
	\definecolor{DarkBlue}{cmyk}{0.70,0.60,0.00,0.60}

	\usepackage{ifpdf}	
						
	\ifpdf
		\usepackage{hyperref}
		\hypersetup{
			pdfpagelayout=TwoPageRight,
			pdfborderstyle={/S/U/W 1},
			colorlinks = true,
			linktocpage = true, linkcolor=blue, citecolor=blue, urlcolor=blue, bookmarks,
			pdfauthor={Fries, Kohl-Landgraf, Paffen, Weddigen, Del Re, Schütte, Bacher, Declara, Eichsteller, Weichand and Streubel},
			pdfsubject={Paper},
			pdftitle={Implementing a financial derivative as smart contract},
			pdfstartview=FitH
		}
	\else
		\usepackage{url}
	\fi	

	\newcounter{cpfNumberOfFigures} 
	\newcounter{cpfNumberOfTables} 

\begin{document}

		\subject{}
		\author{
			Christian Fries
			\thanks{DZ Bank AG Deutsche Zentral-Genossenschaftsbank, Platz der Republik, 60325 Frankfurt am Main, Germany}
			\thanks{Department of Mathematics, University of Munich, Theresienstra\ss{}e 39, 80333 München, Germany}
			\and
			Peter Kohl-Landgraf\footnotemark[1]
			\and
			Björn Paffen\footnotemark[1]
			\and
			Stefanie Weddigen\footnotemark[1]
			\and
			Luca Del Re\footnotemark[1]
			\and
			Wilfried Schütte\footnotemark[1]
			\and
			David Bacher\thanks{Bayerische Landesbank, Brienner Straße 18, 80333 München, Germany}
			\and
			Rebecca Declara\footnotemark[3]
			\and
			Daniel Eichsteller\thanks{Cofinpro AG, Untermainkai 27--28, 60329 Frankfurt am Main, Germany}
			\and
			Florian Weichand\footnotemark[4]
			\and
			Michael Streubel\footnotemark[4]
		}
		\title{Implementing a financial derivative as smart contract}
		\subtitle{
			Key concepts, procedural and legal implications, comparison of technological frameworks, prototype description\\[2ex]
			 \centerline{\small Version 1.0.1}
		}
		\date{February 25, 2019}

	\maketitle

	\begin{abstract}
In this note we describe the application of existing smart contract technologies with the aim to construct a new digital representation of a financial derivative contract. We compare several existing DLT based technologies. We provide a detailed description of two separate prototypes which are able to be executed on a centralized and on a DLT platform respectively. Beyond that we highlight some insights on legal aspects as well as on common integration challenges regarding existing process and system landscapes. For a further introductory note and motivation on the theoretical concept we refer to \autocite{FriesKohlLandgrafOBLB}. A very detailed methodological overview of the concept of a smart derivative contract can be found in \autocite{FriesKohlLandgraf}.

\textbf{Disclaimer: } The views expressed in this work are the personal views of the authors and do not necessarily reflect the views or policies of current or previous employers.
	\end{abstract}
	
	\microtypesetup{protrusion=false}
	\tableofcontents
	\microtypesetup{protrusion=true}
	
	\clearpage
	\section{Introduction}
	
	\subsection{The OTC derivatives market today}
	Financial derivative products are important instruments in risk management within the financial industry. Several forms exist and among those bilaterally traded so-called over-the- counter (OTC) derivatives play a significant role. According to recent statistics \parencite{Bis} the gross market value of interest rate and foreign exchange OTC derivatives amounted to over 10 trillion USD by the end of 2017. Due to their bilateral nature they introduce own contractual risks – especially so-called counterparty credit risk concerning a default and rating migration of the counterparty. An efficient approach to fully remove this risk by construction is still missing. Due to large derivative portfolios and historically grown trade life cycle processing infrastructures procedural inefficiencies exist. These inefficiencies are the reason for the retention of other forms of risk, such as settlement risk.
	
	\subsection{Insufficient process definitions}
	We observe that a derivative contract is defined on different levels of precision, especially when it comes to procedural aspects. A bilateral derivative contract usually has to put its future cash flow structure into writing very rigorously. On the other hand, aspects which are more relevant from a procedural perspective like how to determine a net present value, collateral processes and margining, booking procedure and handling of an early termination are usually not defined on a comparably precise level.
This leads to several uncertainties and inefficiencies during a derivative’s life cycle, which in turn introduce complex processes and risks. For example, there is a dispute resolution procedure \parencite[see Article 15 of][]{eu2013_149} to resolve valuation differences which are required for determining the correct collateralization amount. Redundant booking transactions occur because margin calls and derivative cash flows are settled separately.
Most importantly those insufficient process definitions especially prevent the process infrastructure to evolve towards a more efficient format.

	\subsection{The notion of smart contracts}
	Smart contracts are computer programs. They are decidedly designed to digitally represent a legally binding contract with the aim to support the accurate processing of the contract terms over its life cycle in an automated and standardized way. We emphasize two major advantages in the following sections.
	
	\subsubsection{Deterministic contract definition}
	A legal contract which is represented by a computer program provides a high level of precision. When that program is executed it can only take states or act on events which have been predefined in the code. An event or contract state which is not exactly defined in the program cannot exist in the digital representation. When correctly conceived a smart contract cannot end up in an undefined state. What appears to be the main advantage also turns out to be the main challenge. Before handing control over to a program one has to make sure in advance that the designated piece of software exactly behaves the way the contract is initiated. This includes a complete understanding of every possible state that such kind of computer program can reach and whether all those states are desired by the legally binding contract design.
	
	\subsubsection{Enforceability}
	A computer program reacts in a deterministic manner based on predefined events only and switches between predefined outcome states. It also will handle critical events according to how it is defined. This might be a helpful feature compared to the “real world” as contracts according to current standards may run into an undefined process state. In such a situation both contract parties may disagree on the exact contract state, the final outcome in critical events is uncertain since it depends on human interaction and agreements. With new smart contract standards, also the outcome of critical events becomes fully deterministic and predictable.
From a legal perspective this may raise another challenge whether software can enforce a legally binding transaction even in critical cases – e.g. the question whether in case of a looming bankruptcy an automated transaction triggered by the software will also be legally enforceable afterwards.
However, with regard to financial derivatives – as the contractual and procedural worlds are partly separated and the explicit handling of critical events is complex – e.g. in case of disagreement on accurate collateral amounts – the application of above features can be seen as advantageous.

	\subsection{The advantage of decentralization}
As smart contracts are computer programs, they have to be executed in a specific infrastructural environment. It is appealing to let a third trusted agent act as an independent instance, which controls the correct execution of the contract according to the specified algorithm. An important premise is that both parties need to trust a third party as the concept heavily relies on its operational efficiency. Problems arise if the trusted party does not behave as expected. Furthermore, a third party can produce mismatches in data and therefore separate validation and reconciliation is required for the contributing parties.
In a decentralized version, the position of a central party will be replaced by a network formed by the contributing parties. Over its lifetime, the defined smart contract functionality changes contract states. Any migration to a new contract state triggers all participating parties in the network to verify and validate the resulting transactions in terms of correctness, so in the end that transaction is market as approved by the entire network. This approach requires one immutable shared data source to facilitate successful validation and consensus on the current network state. This fundamental feature is provided by the concept of distributed ledger technology (DLT).
Some distributed ledger-based systems also support the execution of smart contracts. An important feature is that a specific outcome of a contract – e.g. a new state and/or the execution of a specific transaction – needs to be validated by all contributing network nodes. In turn this implies that the operational risk posed by the concept of one central instance can be resolved with a distributed validation processes across multiple parties.
A third alternative can be a hybrid version of both. For some use cases where an instance of trust cannot be omitted a third party is left in place. For example, in case of a permissioned network, the role of a central permission service is required. On the other hand, decentralization of processes which bear operational risks or require high reconciliation effort seems attractive and efficient.

	\subsection{Excursion: Central clearing}
Lately, the financial industry has highly embraced the concept of risk mitigation through centralization and risk sharing. The European Market Infrastructure Regulation (EMIR) directive has enforced a central clearing obligation for interest rate swaps \parencite{Esma} and other standardized OTC derivatives. Since 2016, financial counterparties are not allowed to process an IRS’s life cycle bilaterally and instead are forced to contract over a third institution – a so-called clearing house or central counterparty (CCP). The rulebook of a CCP, which also defines cash flow processing, market value determination and default resolution, can be seen as an approach to fully specify the trade life cycle events.
A question increasingly discussed in the industry is whether this level of centralization is sensible.4 The clearing house itself is exposed to the risk of insolvency of one of its clearing members and can get into serious difficulties depending on the portfolio volume and position of the defaulting party. So-called default resolution procedures, which rely on a joint resolution procedure including all remaining clearing members, are put into place as risk mitigation strategies. Based on the fact that the volume of cleared swaps has increased significantly over the last years, the question remains whether one central instance, including the mentioned member resolution, is the best way to manage a default especially in a possible aligned stressed market environment. Efficacy and transparency of the processes in case of a clearing member default and the implementation of the loss-sharing procedure among the remaining clearing members through default funds are controversial – for example see \autocite{ClancyNasdaq}.
In this context, smart contracts and their implementation using modern distributed ledger technologies are a promising alternative to central clearing since a bilateral and decentralized implementation of a derivative’s life cycle has the charming feature of preventing dangerous risk aggregation on central nodes, thus easing systemic risk and sustaining natural competition. However, a trusted third party might still be required providing several centralized services. In the context of financial derivatives that might be services such as market data and valuation services, issuing digital currencies and permissioning.
	
	\clearpage
	\section{A smart derivative contract}
	In the following section, we give a brief outline of major components of a “smarter” implementation of a financial derivative contract. The concept is inspired by the notion of smart contracts and distributed ledger technology and by several preceding works – see \autocite{Morini} for example. Our concept is platform agnostic in the sense that it does not rely on a decentralized infrastructure. Its advantages may be leveraged on a centralized infrastructure as well. For more technical specification and methodological details we refer to \autocite{FriesKohlLandgraf}.

	\subsection{Deterministic contract terms}
	\label{sec:ImplementingSDC:Section2-1}
	In order to implement a smart contract analogue of an OTC derivative a fully deterministic definition of every trade life cycle event and the corresponding process steps is necessary. This will provide full clarity of all process states in advance – including those, which are currently vague. If those contract conditions are furthermore translated into computer code, we obtain a digital replication of the original derivative contract and its associated processes – a smart derivative contract. To achieve this, the current derivative’s legal contract terms have to be extended by the following five additional elements.

	\subsubsection{Predefined market value determination}
	Since the smart derivative contract represents not only a stand-alone derivative contract but also an equivalent to all currently associated standard processes, it has to integrate the margining procedure. The net present value of a derivative contract is central in order to determine the correct collateral procedure or settlement amount at any point during the lifetime of the derivative. Therefore, we postulate that the market value determination has to be defined exactly and unanimously in the contract terms. This postulation implies several aspects that have to be agreed on a very detailed level:
	\begin{enumerate}
		\item market data reference
		\item valuation models and their parametrization
		\item timing of valuation
	\end{enumerate}
	Each of those aspects comes with challenges of its own, which will be discussed later on.

	\subsubsection{Frequent settlement of market value}
	Collateral and derivative cash flows are currently executed in separate processes, which may lead to redundant bookings. To resolve this, we postulate a settlement of the current market value of the underlying derivative, which resets the market value of the overall contract (net position of derivative plus cash) to zero at a daily or at an even higher frequency (we note the analogy to the settle-to-market model of CCPs). We emphasize that the first postulation (market value determination) is a prerequisite for the second. In terms of risk management, this frequent settlement will reduce current counterparty risk to a one-day gap exposure risk.

	\subsubsection{Introduction of margin buffers}
To execute the settlement, it has to be ensured that both counterparties have provided sufficient liquidity beforehand. Counterparties are obliged to post a specific margin buffer amount to dedicated wallets from which the contract can trigger the settlement booking in an automated manner. The margin buffers are pre-defined in the contract terms. One possibility is to determine the amount using a quantile-based approach as proposed in \autocite{FriesKohlLandgraf}, which allows drawing an analogy to current initial margin models and processes. We like to stress that the margin buffer is conceptionally the pre-funding of an expected maximum variation margin, consumed at every settlement. This is different from current initial margins, which are used only in cases where the margining process fails. In case of an insufficient margin balance the smart derivative contract terminates prematurely.

	\subsubsection{Premature termination}
We postulate that the smart derivative contract can terminate itself automatically in certain contract states. More precisely, the termination is triggered by the following two conditions. Either the account lacks the minimum margin balance or the change in market value is larger than the available balance although the minimum buffer has been posted. The latter can happen due to large market moves which are not covered by the predefined buffer amount.

	\subsubsection{Termination fee}
	\label{sec:ImplementingSDC:Section2-1-5}
	Premature termination can be triggered for different reasons. First, a large market move which is not covered by the margin buffer, will terminate the contract. Second, the contract will terminate in case, one contractual party is not able to post sufficient liquidity to the margin wallet because of a credit event. Third, one party can decide on purpose not to post the required margin amount to the contract wallet. Whereas the first two are features and, at least the second one, even desirable features of the smart derivative contract, the third one is undesirable and a contract feature has to be implemented to avoid it. Any contract party can empty their contract wallet at any given time during the lifetime of the contract, resulting in a willful termination of the contract on the next settlement time (on the same day). Thus, the automatic termination feature turns out to be an American Option. In order to remove the incentive of terminating a contract with a negative market value, a predefined termination fee is supposed to
	\begin{enumerate}
		\item make it economically inefficient to willfully terminate the contract, and
		\item cover costs incurred by the other party in need of a replacement contract.
	\end{enumerate}
	Both contractual counterparties at the inception of the contract post the termination fee to separate segregated accounts / wallets. When the contract terminates due to lacking liquidity in one counterparty’s wallet, the contract automatically and instantaneously transfers the termination fee to the other counterparty at the next settlement time.

	\subsection{Putting the terms into chronological order}
\begin{figure}[hbtp]
		\begin{center}
			\includegraphics[scale=0.66]{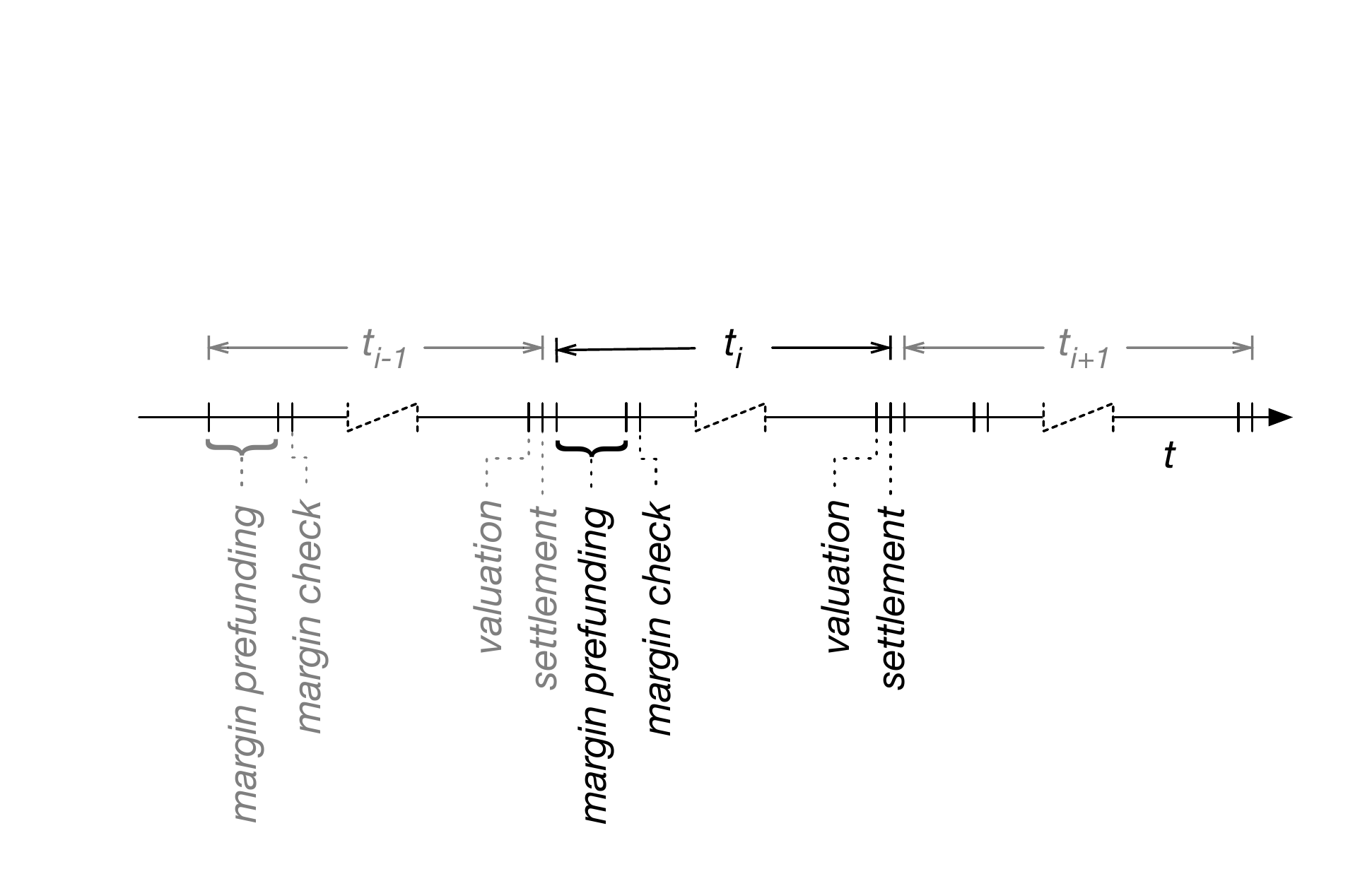}
			\caption[
			]{
				The events of a smart derivative contract.
			}
			\label{fig:ImplementingSDC:smartDerivativeContractTimes}
		\end{center}
		\addtocounter{cpfNumberOfFigures}{1}
	\end{figure}

	The terms defined in the precedent five postulates need to be put into a clearly defined chronological order to implement the desired behavior from a procedural perspective. There is a subtle requirement on the timing: In order to avoid that a counterparty may use information on an upcoming settlement amount, it is required that adjustments to the margin buffers are allowed only at a short period right after a settlement. Since after settlement, the market value of the contract is 0, this ensures that a counterparty cannot use an adjustment of the margin buffer in its own favour. Figure~\ref{fig:ImplementingSDC:smartDerivativeContractTimes} illustrates a time line with the events of a smart derivative contract: margin prefunding, margin check, valuation and settlement, where margin prefunding occurs close to the previous settlement.

	\subsection{Requirements of a smart derivative contact}
	\label{sec:ImplementingSDC:Section2-3}
	The technical and procedural implementation of the given five postulates on a given infrastructure has to meet the following requirements.
	\begin{description}
		\item[Margin wallet] An account to handle frequent contract settlement (this may be an escrow account or a special wallet (see below). Segregated margin accounts are needed to lock away the posted margins until the next settlement. Rebalancing of the margin wallets through the contractual counterparties is allowed in predefined periods. The counterparties cannot access their margin wallets outside these predefined rebalancing periods.
		\item[Cash on ledger] To fully conceive automation and enforceability the contract has to be enabled to book settlement amounts and termination fees autonomously. This can be put into force by introducing digital tokens which can be transferred automatically between wallets. Existing crypto-currencies are inappropriate in this context due to the exchange risk with respect to the contract currency. Therefore, this is a prominent use-case for a one-to-one digital representation of a main existing fiat currency, a so-called stable coin. The installation of a central agent that issues the stable coin and guarantees its conversion between digital coins and fiat money at any time builds trust in the digital currency.
		\item[Central valuation instance] For unique market value determination either of the following is needed:
			\begin{enumerate}[a.]
				\item An independent and trusted source for the valuation or
				\item an independent and trusted source for market data and the ability to perform complex valuation.
			\end{enumerate}
		For valuation, open-source valuation libraries and their routines could be used since they are readily available to both parties. Alternatively, proprietary (to one counterparty), shared or trusted third party valuation engines can be integrated with the contract’s infrastructure. To this end, it might be sufficient to reference a specific version of source code or binaries in the implementation of the Smart Derivative Contract, for example by agreeing upon the hash of a specific version.
		\item[Immutable market data reference] The frequent and automated determination of the definite market value of the contract through the central valuation instance upon which actual margin payments are automatically executed presupposes that market data in high quality is readily available to the contract and/or its valuation instance. Two possible options exist:
		\begin{enumerate}[a.]
			\item Market data is made available inside of the contract’s infrastructure (e.g. written directly to the DLT on which the contract lives) and can be accessed by the central valuation routine
			\item Market data is fed to the contract or its valuation routine from a trusted source outside of the contract’s infrastructure (e.g. a secure shared database)
		\end{enumerate}
		In any case, the market data has to be transparent to all contractual parties and immutable with sufficient historisation routines (e.g. for validation purposes of the central valuation routine or backtesting). The issue of how to ensure that the market data is of sufficient quality, how to proceed with payments that were triggered or, in the worst-case, contracts that were terminated due to corrupt market data, persists.
		\item[Timing and scheduling] The correct procedural realization of the Smart Derivative Contract critically depends on the scheduling and timing of the single contract events as displayed in figure (Ref to process flow diagram). Thus, the infrastructure on which the contract is implemented has to be equipped with a scheduling service and it has to be clear to all contractual parties at which exact time a certain contract event did occur.
		\item[Premature termination] A possibility to prematurely and automatically terminate/suspend the contract under predefined conditions. Termination must be enforced automatically even without having an explicit approval from one or even both counterparties. This can be especially challenging from a legal perspective in case of bankruptcy of one smart contract counterparty.
		\item[Finality of every contract state] The algorithm is not allowed to reach a state where a final validation across several participants cannot be reached. It has to be exactly pre-defined which state needs validation and by whom.
		\item[Privacy]
			\begin{enumerate}[a.]
				\item The contract terms are only transparent to a very limited group of participants, most probably only to the individual contractual parties.
				\item The cash flows do not disclose information about the contract or the counterparties outside a limited group.
			\end{enumerate}
	\end{description}
		
	In Section~\ref{sec:ImplementingSDC:Section5} we will discuss how different technologies cope with these requirements.

	\clearpage
	\section{Integration into a bank’s infrastructure}
	Above contract features are bold as they require severe changes in current processing landscapes of a financial institution. Implementation of the described concept raises several challenges, especially when it comes to the question of how to integrate that concept into existing infrastructures. On the other hand, it is also clear that due to existing trade volumes existing technical landscapes cannot be rebuilt from scratch.

	From a realistic point of view, a very first integration approach of such a new concept should be done in a way that existing infrastructure is modified only at a minimal level and most of the existing toolset (e.g. trading systems) is used to the maximum possible extent.
	
	\subsection{Minimal integration effort}
	There are several reasons that will not allow the smart derivative contract to exist on a distributed ledger alone and invoke the necessity for it to also be represented in the institution’s front-to-back (FtB) system. A multitude of downstream processes and reports exist in financial institutions that rely on the input from the FtB system. According to the minimal integration effort hypothesis, these should not be directly affected by the new digital representation of derivative contracts, making a representation of the smart derivative contract in the FtB system mandatory.

	The following are some of the most crucial aspects that need to be considered in order to successfully integrate a smart derivative contract as a new product into the existing infrastructure of a financial institution.
	\begin{description}
		\item[Financial accounting] Balance sheet reporting is based on the FtB system data stemming from its replication of the respective trading system
		\item[Regulatory requirements] The contract has to meet the applicable regulatory requirements according to the applicable jurisdiction, e.~g. the EMIR transaction reporting obligation \parencite{eu2012_648}
		\item[Risk control] The smart derivative contract has to be monitored in the bank’s existing and approved risk management systems
		\item[Treasury] Liquidity has to be managed, with all cash flows being represented in the FtB system.
	\end{description}
	Furthermore, as long as the concept of cash-on-ledger is unavailable, the smart derivative contract’s cash flow need to be settled in the traditional way via existing payment systems.

	\subsection{Operational handling}
	\label{sec:ImplementingSDC:Section3-2}
	
	\subsubsection{System replication}
	To integrate a smart derivative contract in a standard trading system would require the following two components: A package consisting of two legs: The first leg will contain the underlying swap product (e.g. a plain vanilla interest rate swap) and the second leg will contain a cash account into which the reset amounts are booked. The following example illustrates the replication concept:
	\begin{itemize}
		\item Swap has a positive market value for counterparty A
		
		\item Counterparty B pays the reset value to counterparty A
		
		\item Reset value will be stored as a liability in the cash component (the cash component
becomes a negative value)

		\item The cash component will offset the net present value of the derivative
	\end{itemize}
	Furthermore, it is important that the representation of the smart derivative contract in the bank’s FtB systems is such that cash flows from the underlying trade are not triggered in the traditional payment systems. Also, the underlying trade cannot be part of the traditional margining process, which has to be ensured by a suitable setup in the FtB systems of the package of trades of which the internal representation of the smart derivative consists. Both aspects are fully covered by the smart contract instance.

	\subsubsection{Valuation}
	Unlike traditional OTC derivatives, the valuation of the smart derivative contract is not based on each counterparty’s internal valuation model but relies on an external valuation source, a so-called oracle. Thus, the underlying derivative’s net present value is exogenously determined and contractually accepted by both parties. As a result of the oracle’s valuation, the reset value is exchanged between the counterparties on each settlement date.

	Right after a smart derivative contracts settlement its economic value is zero, since the counterparty could withdraw any positive amount from the margin buffers and let the contract terminate without further liabilities. However, the respective counterparty’s internal valuation determining the value that is used for the balance sheet and regulatory reporting purposes might at a different valuation time point and with different market data usage. That existing gap is not mitigated through the contract’s settlement routine. The valuation difference particularly depends on the time gap between valuation by the oracle and settlement via the counterparty’s traditional payment systems. Therefore, both time points are supposed to be as close as possible. Alternatively, internal valuation routines can be skipped overall with a further integration step in which the smart derivative contract feeds the current product valuation to the bank’s internal systems at each settlement time and the internal trade representation is in fact nothing more than an empty shell. In this case, balance sheet valuation and “execution valuation” by the oracle, which determines the trade cash flows, are in line and there is no additional gap risk. A suitable workaround for risk control processes (e.g. value-at-risk calculations) and systems (e.g. limit and counterparty risk systems) has to be implemented.

	\subsection{Procedural adjustments}
We describe several aspects regarding existing variation and initial margin processes.

	\subsubsection{Variation margin and initial margin}
	The smart derivative contract is subject to the existing Master Agreement between the parties and not to the clearing obligation. Thus, pursuant to Article 11 (3) of EMIR \parencite[for details s.][]{Bafin}, it must be collateralized. By its very construction, there is supposed to be no exchange of variation margin for this product. In order for the smart derivative contract to be handled correctly in the collateral management system (CMS), it is necessary for the package (or unit) to have a market value of (close to) zero. In conclusion, the smart derivative contract changes the nature of the underlying derivative product from so-called \emph{collateralized to market} (CTM) to \emph{settled to market} (STM, \autocite[s.][]{clarus}) and substitutes the variation margin process (VM) with daily (or even more frequent) settlement procedures realizing the market value. Since the economics and the valuation of the underlying derivative product should not change by this procedural aspect, it should be valued in the same way as it was under the collateralized-to-market. For that reason, so-called \emph{price alignment interest} (PAI, \autocite[s.][]{IsdaCtm}) – i.e. overnight funding costs of fictitious collateral – will still be taken into account (e.g. through OIS discounting).

	Regulatory bilateral initial margin (IM, \autocite{eu2016_2251}) requirements will apply also for a smart derivative contract, hence initial margin amounts based on the risk profile of the underlying derivative product currently have to be exchanged between counterparties. It is open to further research and to a possible adjusted regulatory treatment whether the contractually defined termination fee (Section~\ref{sec:ImplementingSDC:Section2-1-5}) may already be treated as an initial margin.

	\subsubsection{Payment and reconciliation processes}
In a first step, although cash-on-ledger is still unavailable and payments might have to be processed through the existing payment processes of the bank, the following process simplifications can still be achieved:
		\begin{itemize}
			\item Changes in market value and derivatives product cash flows are netted to form the reset amount
			\item The transition from CTM to STM processing omits the duplication of the payment process for product and collateral cash flows
			\item Due to the contractually agreed external valuation, market value reconciliation between counterparties is not necessary anymore. Time consuming and complex dispute processes are eliminated
In a second step after the introduction of cash-on-ledger the procedural efficiency of a smart derivative contract will increase further.
			\item Automatic and possibly highly frequent intraday payment process via digital accounts, i.e. wallets
			\item The only payment process necessary via traditional methods has to be done by a treasury department refilling or withdrawing from the wallets according to previous and anticipated market movements, covering the margin buffer amount at any time
		\end{itemize}
    \subsubsection{Product innovation and clearing obligation}
The valuation standardization that is introduced through smart derivative contracts alters the process chain towards a higher degree of automation. It is beneficial for both counterparties as transaction, settlement and collateralization processes can be transformed into a far more efficient setup. By eliminating parts of the traditional process chain (i.e. market value reconciliation, variation margin process), operational risks as well as process complexities are significantly reduced. Due to its optional component – i.e. the inherent daily termination option for both counterparties –, the current legal assessment (which is yet to be confirmed by the regulators) is that smart derivative contracts do not fall under clearing obligations. Thus, the new concept allows classical derivative products to be traded in a bilateral way which otherwise would have to be cleared. This is favorable from the perspective of systemic risk, opposing the current concentration of risks with central counterparties.

	\subsubsection{Documentation and regulatory acceptance}
To integrate the new features of the smart derivative contract they need to be embedded in the existing legal framework. This can be legally designed in form of a new \emph{master confirmation agreement} (MCA) and by adding additional contract terms to OTC trade confirmation. We discuss further legal aspects in the next section.

	\clearpage
	\section{Aspects from a legal perspective}
	In this section aspects and implications from a legal perspective are discussed. Several aspects of the smart derivative contract are considered with regard to two other recent publications \autocite{IsdaConceptToConstruction} and \autocite{Linklaters}.

	\subsection{The notion of \enquote{code is law}}
Smart contracts may have the potential to increase efficiency, timing and performance by automating and processing contract terms in computer code. There is a legal discussion whether contractual provisions solely formally expressed by computer code are able to have a legally binding and effective character. This is especially important if the contractual conditions are expressed completely and solely in terms of computer code – code is law in that case. The publication \autocite{Linklaters} introduces the notion of a smart legal contract. Notably, basic legal issues as the identification of the contracting parties, interpretation of legal conditions and relevant governing law may raise legal concerns under these circumstances.

	\subsubsection{Limitations from a legal perspective}
The question is to what extent computer code can properly deal with any kind of legal clauses. Computer code may be well suited to capture contractual states that are based on logical expressions. They may not be able to capture legal facts or issues based on legal notions such as rationality or conscience. Difficulties arise when it comes to discretions that are beyond a clearly defined framework. Computer code is not able to express some present contractual conditions in the context of derivatives, e.g. the following legal expression “The price has to be adjusted in an economically reasonable manner”. This condition depends on a future not yet known meaning and exercise of conscience, rationality and scope of discretion. In those cases, a codification is not possible due to a lack of logical relevance. An attempt to do so would create the risk of deviation between the legal meaning of the original contractual regulation and its interpretation by the code. We discuss two concrete events which are critical from that view but may be solved by implementing our proposed concept.

	\subsubsection{Example: Dispute procedure}
If possible, the contract’s terms should be designed to avoid those states where code cannot exactly describe the corresponding legal state. If this is not possible, the states which can be formally expressed need to be identified. \autocite{Linklaters} makes a distinction in operational clauses which are able to be expressed by logical expressions and non-operational clauses. Among others \autocite{Linklaters} lists one example of a non-operational state: disputes. With regard to our concept above we transformed that non-operational state into operational one (see Section~\ref{sec:ImplementingSDC:Section3-2}). For a smart derivative contract the exact determination of the net present value (in terms of market data, valuation model and its parametrization) becomes part of the contract. In the legal terms this can be done by referencing a dedicated version of the referenced software package as well as referencing concrete market data source and associated market data symbols.

	\subsubsection{Example: Early termination}
The publication \autocite{IsdaConceptToConstruction} highlights the point that a smart contract protocol needs to offer a certain possibility to terminate its automated performance. Such kind of early suspension right is seen critical. If a financial derivative contract offered both parties a right to terminate the contract at every point during its life cycle this will nullify its economic added value - namely reduction or elimination of market risk. The issue of early termination is left open in \autocite{IsdaLegal} for further research. Within our concept we introduced the precise possibility of a premature termination and defined that event and its causing states at a very precise level. Our Smart Derivative Contract may terminate early if a margin account does not show sufficient balance or an automated settlement cannot be executed. However, due to the pre-funded termination fee (Section~\ref{sec:ImplementingSDC:Section2-3}) the termination event is made rare as the causing party loses that amount when terminating the contract. There may be cases where an early termination may show to be beneficial in retrospective namely when the lost termination fee is less than the resulting owed settlement amount. Due to the fixed process schedule this possible gain cannot be determined in advance. On the other hand, an unexpected early termination of the contract due to unexpected large market moves can be contractually prevented by agreeing on sufficient respectively conservative margin buffer amounts. See \autocite{FriesKohlLandgraf} for more details.

	\subsubsection{Aligning code and documentation}
With regard to the difficulties above we follow a suggestion made in \autocite{Linklaters}. For the time being, it seems better to implement a smart contract by technical means but to continue to map the legal documentation in an adapted, written documentation format. This leaves the challenge to perfectly synchronize the legal text with corresponding computer code. The resulting software provides mechanisms for an automated execution and enforcement of the contract legal terms. That design comes close to the concept of an “external smart legal contract” as proposed in \autocite{Linklaters}. A Smart Derivative Contract as described here will then be documented in an existing Master Agreement (e.g. ISDA Master Agreement or “Deutscher Rahmenvertrag DRV”) and confirmed by means of a written trade confirmation.

	\subsection{Legal documentation}
	\subsubsection{Master confirmation agreement}
	With regard to the new contract terms a master confirmation agreement can be introduced to increase efficiencies also on the documentation side. That agreement covers and unifies any additional clauses or mechanisms that are specific to the smart contract functionality. That agreement may be placed hierarchically between an existing master agreement and a single trade confirmation. This enables a legal department to keep a simple and precise arranged trade confirmation format for the single transaction itself. With regard to the underlying derivative transaction, the legal documentation process is hardly modified.

	\subsubsection{Parallel existence of documentation and software}
	In a first step a smart derivative contract will be represented by computer code and is therefore limited to a technical representation and automation for a relevant single trade transaction. It will be controlled by the implemented software code an executed in a suited technological infrastructure as discussed above.

	Any arising legal state from the contract terminology (e.g. termination event) needs to be documented in the new agreement format introduced above.

	\subsection{Other legal aspects}
	\subsubsection{The smart derivative contract's termination fee}
	The smart derivative contract’s termination fee may be seen critical from a legal point of view. If termination is following a counterparty’s default, the payment of the termination fee to the surviving counterparty may be conflicting with existing legal regulations.

	The issue can be solved by transferring the ownership of the termination fee to a third party, removing any claim to it. The contract then becomes similar to a credit derivative – triggering a payment upon a default.
	
	\clearpage
	\section{Available technological frameworks}
	\label{sec:ImplementingSDC:Section5}
	In this section, we present several options in the technical implementation of a smart derivative contract and compare available technologies. A first distinction can be made between centralized and decentralized solutions. The basic idea of the latter – summarized under the term distributed ledger technology (DLT) – seems particularly attractive: As we have seen, in the current standard, derivative transactions are booked and valued in separate private trading and accounting systems for each participating party. That implies a high reconciliation effort for each party. In contrast, within a DLT a unique and public ledger exists which results in a unique definition and understanding of transactions. The above concept in consideration poses three major requirements for a technical solution. In our context, this feature can be used to simplify the processes involved in derivative transactions as shown below.
	\begin{itemize}
		\item Autonomous execution of the process without interference from the counterparties
		\item Interaction with external components such as the valuation oracle
		\item Booking of margins based on rules defined by the process
		\item Termination and termination handling
	\end{itemize}
	In Sections~\ref{sec:ImplementingSDC:prototype1CentralizedSolution} and~\ref{sec:ImplementingSDC:prototype2DLTSolution} we describe two explored technical approaches. The first prototype is built on a centralized infrastructure on a spring state machine. The second prototype is constructed to be executed on a distributed ledger platform.

	\subsection{Common elements: timing events and oracles}
	\label{sec:ImplementingSDC:Section5-1}
	While at first the two approaches appear to be fundamentally different, it turns out that they share many elements. We like to highlight some of these common elements:

	\subsubsection{Timing and event triggers}
	The smart derivative contract requires a trusted and common understanding of time (e.g., to trigger the settlement event). There are multiple ways to implement these triggers, for example:
	\begin{itemize}
		\item Active event triggering: A trusted third party is triggering the events on predefined times. This solution has the advantage that the third party can perform some pre-checks, e.g., the availability of market data.
		\item Passive event triggering: The participating counterparties trigger the events at will (e.g. depositing margins, requesting settlement) and the contract merely checks the admissibility of the event. This solution has the advantage that it can be implemented without a third party, as long as the contract has a reliable understanding of time.
	\end{itemize}

	\subsubsection{Valuation oracle}
	The smart derivative contract requires a trusted and common understanding of the net margin derived from the agreed valuation model.
	
	As shown in \autocite{FriesKohlLandgraf}, the net cash flow that has to be transferred upon a settlement at time \(t_{i+1}\) given that the last settlement occurred in \(t_i\) can be expressed as a function
	\begin{align*}
		F(t_i,t_{i+1}) = V(t_{i+1},M(t_{i+1}))-V(t_{i+1},M(t_i))\text{,}
	\end{align*}
	where \(V(t,M(s))\) determines the time \(t\) valuation of the underlying product’s future cash flows and \(M(s)\) is the valuation model calibrated to the time \(s\) market data.

	While the formula clearly requires a derivation (it can be found in \autocite{FriesKohlLandgraf}), it can be summarized and motivated shortly: since the contract performs a (virtual) netting of cash flows and collateral flows all variation margin comes the change in market value induces by the change of the market data \(M(t_i)\rightarrow M(t_{i+1})\).
	
	The valuation oracle is then provided by a service implementing the interface, here stated in Java:
	\begin{lstlisting}[language=Java]
interface SmartDerivativeContractMarginOracle {

	Amount getMargin(LocalDateTime periodStart, LocalDateTime periodEnd);

}
	\end{lstlisting}
	Since market data has to be provided by a trusted source, we may also consider the valuation oracle as being provided by a trusted source, independent of the network solution chosen. Our two prototypical solutions discussed in Sections~\ref{sec:ImplementingSDC:prototype1CentralizedSolution} and~\ref{sec:ImplementingSDC:prototype2DLTSolution} use the same oracle service.

	\subsection{Centralized solution}
	\label{sec:ImplementingSDC:prototype1CentralizedSolution}
	
	Our centralized prototype implementation of the smart derivative contract is a Java application based on Spring State Machine. It can run on a single server and can also be replicated on a server cluster.

	\subsubsection{Concept}
	Spring State Machine is a framework designed to use traditional state machine concepts within Spring applications.
	
	The structure of the smart derivative contract is a sequence of states, which are reached in accordance with a prescribed schedule and following contractually defined rules. By adopting the state machine pattern, the flow of events, i.e. transition into the predefined contract states, is easily split into small manageable tasks which provide loose coupling and easy modularization. The flow implementation mirrors all the rules which are the constituents of a smart derivative contract as described in the previous sections.

	\subsubsection{Operational model}
	There are multiple, largely equivalent representations of the smart derivative contract process in terms of state diagrams.
	
\begin{figure}[hbtp]
		\begin{center}
			\includegraphics[scale=0.55]{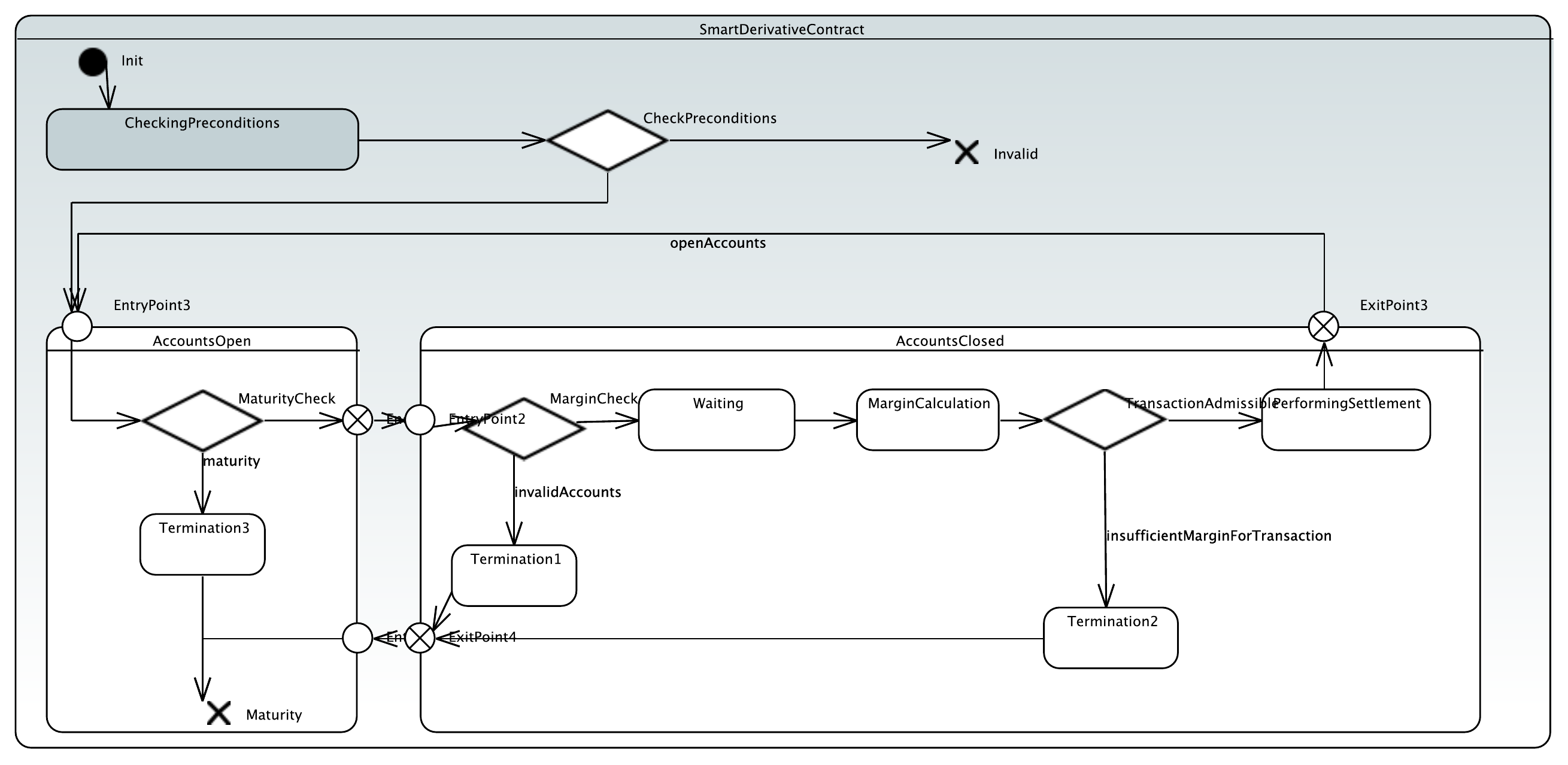}
			\caption[
			]{
				State machine representation of the smart derivative contract.
			}
			\label{fig:ImplementingSDC:stateMachineRepresentationOfTheSDC}
		\end{center}
		\addtocounter{cpfNumberOfFigures}{1}
	\end{figure}

	In Figure~\ref{fig:ImplementingSDC:stateMachineRepresentationOfTheSDC} we present one possible model as an UML state diagram.

	We give a short description of the states: after checking the preconditions of the contract, e.g., the presents of the termination fee (Section~\ref{sec:ImplementingSDC:Section2-1-5}), the contract distinguishes two states: \texttt{AccountsOpen}: the counterparties are allows to post or withdraw amounts from the wallets, \texttt{AccountsClosed}: the counterparties are not allowed to access the wallets. When accounts are closed, the contract processes several sub-states: The contract checks the margin buffers (\texttt{MarginCheck}) and triggers a termination if margin pre-funding is insufficient (\texttt{Termination1}). If margins buffers are sufficient, the contract enters a waiting state until valuation and settlement are due. After the valuation is available the margin call is calculated (\texttt{MarginCalculation}). If the margin buffers are sufficient, a settlement occurs and the cycle restarts, otherwise the contract terminates (\texttt{Termination2}), which includes a possible partial settlement and the transfer of the termination fee. The contract also checks for maturity, which triggers a regular termination (\texttt{Termination3}), where the individual termination fees are posted back to both counterparties’ wallets. After a termination accounts are open for withdrawal.

	\subsubsection{Timed triggers and alternative state diagram}
\begin{figure}[hbtp]
		\begin{center}
			\includegraphics[scale=0.45]{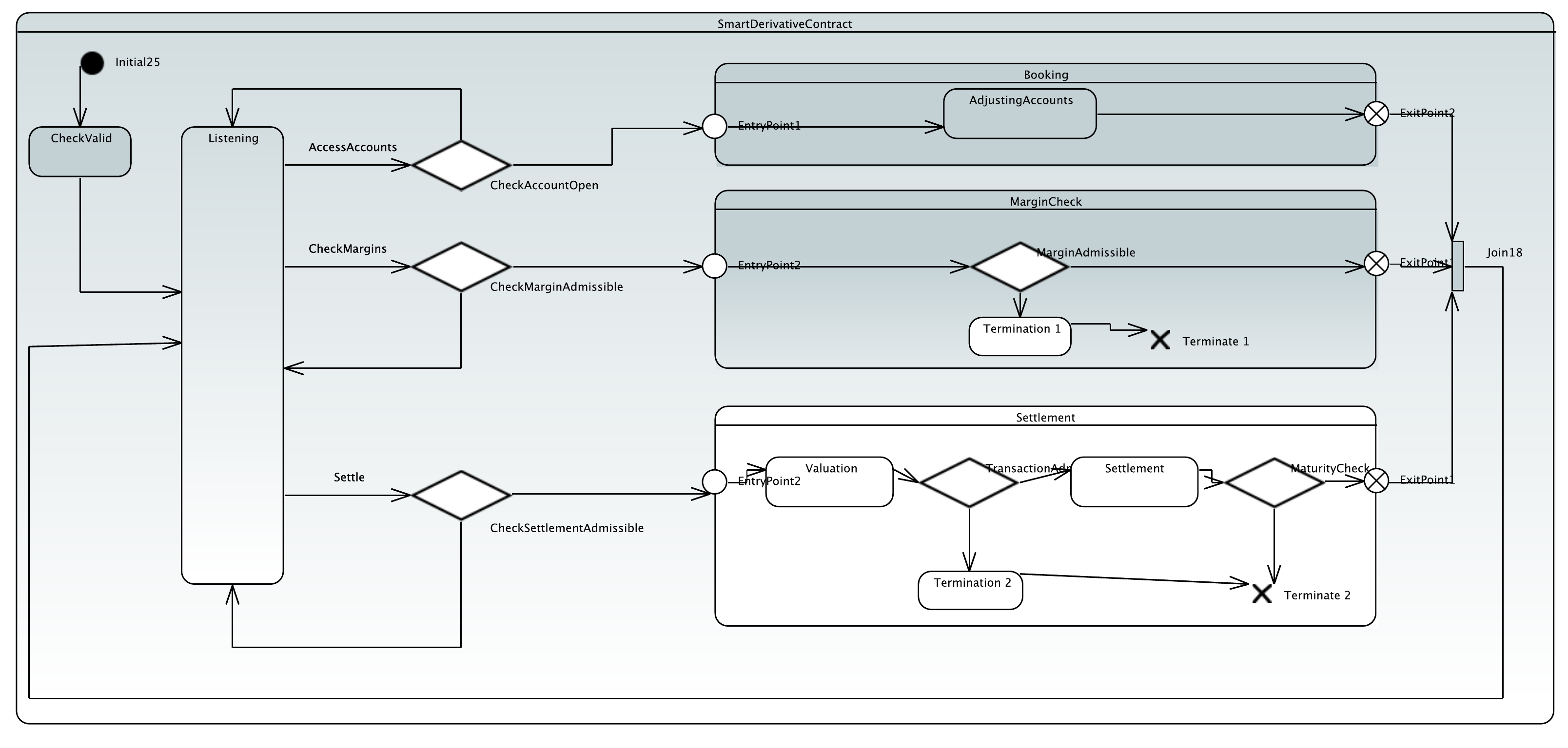}
			\caption[
			]{
				State machine representation with event triggers.
			}
			\label{fig:ImplementingSDC:stateMachineRepresentationWithEventTriggers}
		\end{center}
		\addtocounter{cpfNumberOfFigures}{1}
	\end{figure}

	The state machine diagram in Figure~\ref{fig:ImplementingSDC:stateMachineRepresentationOfTheSDC} leaves it open how a state transition is triggered. In the specification of the smart derivative contract the state transitions (e.g. settlement) occur at pre-defined times. In Figure~\ref{fig:ImplementingSDC:stateMachineRepresentationWithEventTriggers} we depict the states of the smart derivative contract (right side of the figures), which are triggered by some event loop (left side of the figures). See Section~\ref{sec:ImplementingSDC:Section5-1} for alternative to trigger these events.

	\subsubsection{Implementation frameworks}
	Our prototype implementation operates on a conventional relational database which hosts the accounts of the counterparties as well as the contracts with all their relevant properties. The system interoperates with an external (REST) service, which provides the market valuation needed for the margin calculation and settlement. This service could be set up to communicate with a commercial market data service. Such a service is sometimes oracle, e.g. in the context of distributed ledger technologies.
	
	The state machine application has several modes of operation:
	\begin{enumerate}
		\item Its operation is driven by a regular time schedule and a predefined timestamp pattern triggers state transitions and events.
		\item Its operation is driven by an irregular time pattern prescribed by external sources, namely contractual rules and triggers defined in the contract between the counterparties.
		\item Its operation is driven by “state machine drivers” which are time pattern agnostic and prescribe the sequence of events executed on the machine. In this mode, the machine can also be used as a simulation machine running the defined event sequence in as many loops as required by the contract.
	\end{enumerate}
The underlying database hosts the counterparty accounts as well as the contracts that the machine should process.

	\subsubsection{Critical hypotheses}
	\paragraph{Infrastructure} The application code can run safely in a suitable environment. That is, some central and trusted party functions as the host of the required hardware. It controls and manages all technical aspects of the entire underlying infrastructure and is responsible for the stability of the service.

	\paragraph{Security and permission} The state machine application can be sufficiently secured according to existing banking regulations but also in terms of method and data access permissions, i.e. only privileged roles can access code and/or counterparty and contract data. In practice, the central party that provides the technical infrastructure will also manage and control data security. Also, the central trusted party has to implement a permission service to ensure that all participating parties are viable and trusted counterparties.

	\paragraph{Interfaces} Communication of the application with the valuation oracle, market data services is set up in a robust and secure way. The application has interfaces with classical bank payment systems (e.g. using SWIFT) which requires massive integration with back-office systems, so that all payments are triggered and executed in an automated way.

	\subsection{Decentralized solution}
	\label{sec:ImplementingSDC:prototype2DLTSolution}
	
	A decentralized solution is a distributed application, which is executed by every participating counterparty. In this section, we provide a brief overview of existing distributed ledger concepts.

	\subsubsection{Block chain and distributed ledger: Foundations}
	In order to understand the possibilities and limitations of a distributed application or system we introduce their basic principles.
	
	A block chain is a list of records (so-called blocks) which are linked using cryptography. Each block contains a value that represents the previous block. The previous block is identified by a so-called hash, which is transforming data into a string of fixed length using an algorithm. This way, each new block continues the list and therefore links the blocks to a growing chain. If a block needs to be altered, all subsequent blocks need to be changed as well, since they contain a representation of the altered block.
	
	Using this method, a block chain can be used to record transactions securely between two parties in a permanent and almost immutable way. In order to use this principle between different parties, a peer-to-peer network needs to be implemented. Typically, a block chain is run within such a network adhering to a specified protocol. This protocol defines inter-node communication and validation of new blocks.
Validation of blocks is needed because the recording of transactions ideally contains only valid and true facts. A record should not contain false information. Regarding a transfer of money for example, a system should contain a record of correct balances. Validation within these peer-to-peer block chain networks can be achieved in numerous ways.

	\subsubsection{Concept of validation}
	Validation itself is the act of making sure that a transaction is valid. Typically, a block chain or DLT uses signatures to allow participants to validate whether another party has created a certain transaction. If node A signs a transaction, it proposes a new state to the network. Now every other node is able to verify the transaction from node A. If consensus is reached on the validity of the new transaction, it is included into the next block. This way, if a dispute arises later on, every transaction at the root of the dispute can be retraced to find the correct solution by checking signatures.
	
	Apart from confirming the identity of the proposing parties, another part of the validation process consists of verifying whether the proposed transactions are consistent, for example by checking whether the amounts are covered by the wallet balances.
When using smart contracts, validation also needs to be embedded within the underlying protocol by defining certain rules that need to be followed in order to include a transaction into a block. If these rules are not fulfilled, a transaction cannot be part of a block and is therefore denied. These rules could consist of sanity checks like that a function has not been altered or that the contract versions need to match when executed on different nodes. This way, the system ensures that invalid (or malicious) transactions are not included.

	\subsubsection{Concept of consensus}
	Consensus is used to make sure that the state of the chain is valid for all participating nodes. In Bitcoin and Ethereum, new blocks of the chain are generated by so-called miners. There are multiple miners across a network. When a node proposes a new transaction and sends it to the entire network, it inevitably reaches all miners. The miners validate the transaction based on signature and other rules defined in the protocol and then add the transaction to a new block they generate. However, based on latency and other transmission mechanisms the transactions will not reach every miner at the same time. Therefore, each miner might generate a block with different sets of transactions.

	This is where consensus comes into play in order to reach an agreement between the participants and determine which block should be added to the chain of blocks. This is important since the order of transactions determines whether someone did double-spend or not: Having 5 units of money in our system, spending them, receiving the same amount from a second party and spending the same amount again is not the same as spending 5 units twice and then receiving 5, although the end result is the same. An unambiguous timeline of transactions is important and consensus is the mechanism that ensures chronological consistency. There are several ways to reach consensus. Following up three popular consensus models are explained.

	\subsubsection{Concept of \emph{proof of work}}
	\emph{Proof of work} is the consensus mechanism used by Bitcoin and Ethereum and was invented long before them. Basically, proof of work imposes a very difficult and/or expensive operation or action, to disincentivizes misuse of thereby guarded services or actions. For example, if you had to run 5 miles before you were allowed to watch TV, this would be a disincentive to you to watch TV in the first place. You would only watch TV, if you really had to or wanted to and therefore accepted running 5 miles first.

	Using proof of work in the context of block chains, new transactions can only be added after a certain task has been completed. This work typically contains a very difficult cryptographic task, which needs to be solved. Only if the task is completed correctly, a new transaction can be added by the node which solved it. These tasks not only take a rather long time to solve, they also require a lot of economic effort in terms of energy, memory and parallel execution.

	Miners compete against each other to solve the imposed cryptographic task The first miner to solve the task is allowed to add its new block of validated transactions to the chain and in return receives a monetary reward. The other miners are notified and updated with the added blocks, and then restart to build their next blocks. This way, the system tries to ensure that only serious miners that are willing to proof their work and pay the economic price actually participate.

	\subsubsection{Concept of \emph{proof of stake}}
	Since proof of work is very expensive and questionable concerning sustainability, block chain developers came up with other consensus models. One of them is \emph{proof of stake.} Proof of stake basically chooses a miner which is allowed to add the next block based on weighted randomness. The higher the stake of a miner, the higher the probability that their block is selected as the next block.
	
	There are different variants of proof of stake. Some are dependent on the amount of money/coins a miner possesses on the system, others depend on the age of the participants. That is, the system requires a miner to either spend a lot of money or participate for a long time before it is likely to add a block. This way, malicious use is made difficult.

	\subsubsection{Public versus permissioned versus private networks}
	Public networks are block chain networks that are publicly available to anyone who is willing to join, for example the so-called mainnet of Ethereum.

	Permissioned networks are usually managed by one or more central nodes that provide access to the network, for example by using a certain security token, or private-public key pair to authenticate a user. This might result in more public permissioned networks, or more private ones, depending on the strictness of access.

	Private networks are not available publicly and typically require a node within a privately established network (for example via VPN) to participate.

	In the following we look at two existing infrastructures in more detail.

	\subsection{Ethereum infrastructure}
	Ethereum is an open source, public, block chain based distributed system, that allows computing and operation via scripting. The scripts are called smart contracts. They allow a broad variety of custom functions which make Ethereum ideal for complex applications and functionality. It uses Proof of Work for consensus and is soon to be forked to utilize Proof of Stake. Various implementations of Ethereum are publicly available on a couple of different networks, divided in the mainnet and several different testnets.

	\subsubsection{Ethereum nodes}
	Nodes in Ethereum are basically computers or servers that participate in the network. They are either full nodes, which store a full copy of the block chain, or light clients, which store a part of the block chain.
Other participants are also needed: Miners. They create new blocks and therefore mine Ether, Ethereum’s currency, according to the proof of work consensus model.

	\subsubsection{Smart contracts in Ethereum}
	Smart contracts are scripts based on different programming languages. Programmers are able to customize functionality by implementing smart contracts that provide custom functions and fields. Available programming languages include Solidity, Serpent, LLL, Viper and Mutan. The contracts are compiled to EVM bytecode, which can be executed by the Ethereum block chain.

	Since smart contracts are customizable, any programmed logic can be implemented. This opens up the possibility to implement rules that ensure legal requirements or represent a specific process flow. Contracts can store information, provide utilities and information to participants or other contracts or function as a manager for signatures, values and cash (on ledger). Contracts can also be public in order to be utilized between different parties. Any participant is able to deploy a smart contract to the system, however, this also means that certain rules regarding access of the deployed contracts need to be considered.

	When a smart contract is deployed to the system, it is distributed to each node within the system and receives an address (within the chain). In order to access a function of a specific contract, the function caller needs to know the address and use it to call the function. Each node therefore holds a copy of the transaction history (block chain) and a copy of the contract history. Whenever a participant or other a contract wants to access a contract function, this function is run on every participating node, resulting in changes of states that are then validated and included into the next block by miners according to the consensus mechanisms and protocol.

	Since contracts have addresses, they can be addressed directly, resulting in a couple of interesting possibilities:
	\begin{itemize}
		\item Contracts can be referenced (and therefore occur in lists/mappings)
		\item Contracts can hold tokens / coins (since tokens basically use mappings to keep track of
balances)
		\item Contracts can be duplicated and or versioned
	\end{itemize}

	\subsubsection{Decentralized applications (DApps)}
	Decentralized applications (DApps) are applications, that are executed on the Ethereum block chain, and therefore on every participating node. DApps are implemented via smart contracts that connect them to the block chain. A DApp is the sum of all contracts needed to implement the application, as well as other services such as oracles, and frontends.
	
	Since DApps run on every node, execution of DApps is rather expensive compared to traditional centralized server applications. First, they need to be executed multiple times enforcing redundant results. Second, in order to reach redundant results, expensive consensus and validation algorithms are used. Third, storing these results implies having multiple copies on all nodes.

	However, there are also advantages of Decentralized Apps. On central servers, attackers can easily manipulate data once they breach the single server. Since a centralized server is trust based, there is (out of the box) no control mechanism. Within a decentralized system integrity is provided at any time since every change requires consent between at least a majority of nodes. An attacker would need to breach multiple (at least 51\% in most cases) nodes in order to actually be able to manipulate data. Also, downtime of single nodes is easily compensated by other participants (based on the amount of participants).

	\subsection{Quorum infrastructure}
Quorum is an extension to Ethereum that provides additional functionality by altering the protocol. The added features contain alternative consensus models \emph{Raft} \parencite{184040} and \emph{Istanbul Byzantine Fault Tolerance} (IBFT), as well as the addition of private State DBs to handle private transactions between nodes. With these additions, Quorum is well-suited to be used for private and/or permissioned networks.
Basically, all principles of Ethereum apply. However, some additional changes were made to components of the underlying Ethereum protocol. These changes are described shortly in the following sections. The details can be found in the documentation of Quorum.

	\subsubsection{Quorum nodes}
In Quorum, nodes also exist, however, due to different consensus algorithms like Raft \parencite{184040}, miners are not needed. Depending on the consensus model, other roles exist instead like, for example Raft has leaders and followers, whereas IBFT\footnote{IBFT is an Ethereum Improvement Proposal, with many details at \url{https://github.com/ethereum/EIPs/issues/650}} has validators.
	\subsubsection{Private transactions}
	In Ethereum, contracts and the resulting states are stored in a public state database which contains all blocks/states. Quorum adds a second, private state database. Contracts can be marked as private via a transaction parameter privateFor. The payload of transactions concerning private contracts is hashed, which means that only nodes that are specified within the privateFor parameter are able to decrypt the payload and therefore hold the contract code. Other nodes simply ignore these transactions. The private contracts and their resulting states are then stored within the private state database of the participating nodes to prevent dissent with nodes not involved in the private transactions.

	\subsubsection{Private and permissioned networks}
	Additionally, Quorum is designed as a permissioned network. Since permissioning allows control mechanisms that make Proof of Work or Proof of Stake irrelevant, other consensus algorithms can be utilized. Quorum provides a Raft-based consensus as well as a IBFT based consensus.

	\clearpage
	\section{Description of a DLT based prototype}
	The following sections describe the architecture and different building blocks developed to fully implement a functioning Smart Derivative Contract with all involved participants on the Ethereum/Quorum \autocite{Quorum} infrastructure. This prototype is designed to analyze, identify and test if the explicit requirements from Section~\ref{sec:ImplementingSDC:Section2-1} can be met and to assess the possibilities and limitations of Ethereum/Quorum in this context.

	\subsection{Components and participants}
\begin{figure}[hbtp]
		\begin{center}
			\includegraphics[scale=0.45]{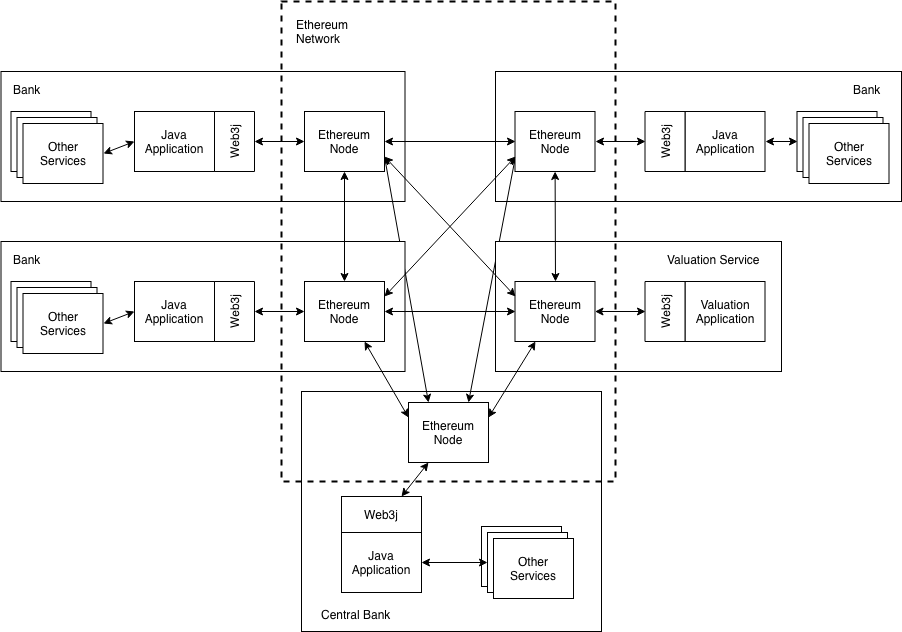}
			\caption[
			]{
				The participants and components of the prototype network.
			}
			\label{fig:ImplementingSDC:participantsOfPrototypeNetwork}
		\end{center}
		\addtocounter{cpfNumberOfFigures}{1}
	\end{figure}

	Figure~\ref{fig:ImplementingSDC:participantsOfPrototypeNetwork} shows the components of and virtual participants in the prototype network.

	The main participants are banks, each running an Ethereum/Quorum node and thus forming the network. Within this network, contracts can be deployed. In order to interact with these contracts, each bank also runs an application serving as interface, for example a Java application that is able to access the bank’s node via web3j. Other services might be added to the context by interaction with the Java application.

	Since valuations are rather complex calculations, execution on the chain would be very expensive. Additionally, the existing algorithms are not written in EVM-interpretable languages but rather in C, C++, Python, Java and Scala. Valuations therefore need to be conveyed to the network from off-chain. A (central) provider for these off-chain valuation services is needed.\footnote{Although in many use cases rather unlikely due to their competitive nature, in pricinple one of the contractual counterparties can take the position of the valuation provider thereby employing his own proprietary valuation algorithms.} Valuation is delivered as “facts” to the network via for example a dedicated valuation oracle. This (central) valuation instance obviously needs to be a trusted source. The valuation instance needs to be able to access valuation algorithms counterparties that the have been agreed on without sharing them with other participating parties. In Figure 5 this instance is called valuation service. It is a participant in the network and also runs a node within the Ethereum/Quorum cluster so that it can access all the contracts that make use of the oracle The Valuation Service node is, as before, accessed via a Java application through web3j. The last participant provide in the designed network is called central bank and acts as the issuing instance of a stable coin on the ledger (cash-on- ledger). That instance issues and controls custom tokens representing real world cash values. It also runs an Ethereum/Quorum node to interact with the network and accesses functions through web3j and a Java application.

	\subsection{Infrastructure: network nodes and contracts}
\begin{figure}[hbtp]
		\begin{center}
			\includegraphics[scale=0.45]{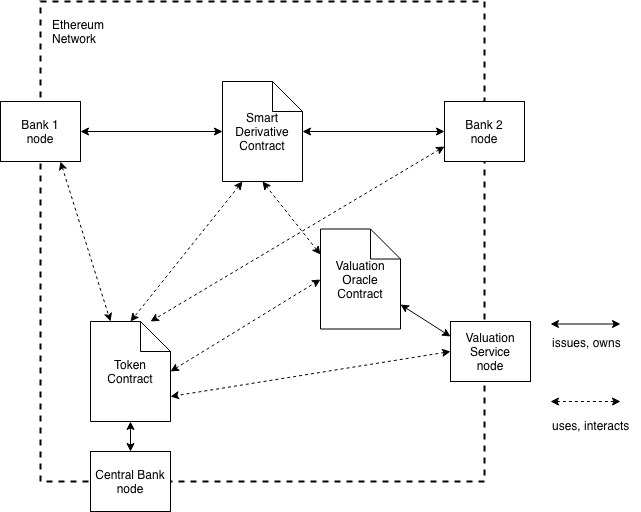}
			\caption[
			]{
				Interplay of the valuation oracle, token oracle, smart derivative contract.
			}
			\label{fig:ImplementingSDC:interplayOfContractAnOracle}
		\end{center}
		\addtocounter{cpfNumberOfFigures}{1}
	\end{figure}

	Figure~\ref{fig:ImplementingSDC:interplayOfContractAnOracle} depicts the contracts which exist on and interact within the network. In order to simplify the image, a minimum amount of nodes is considered. The following three contract types are being implemented:
	\begin{itemize}
		\item Smart Derivative Contract
		\item Valuation Oracle Contract
		\item Token Contract
	\end{itemize}
	
	Smart derivative contracts are smart contracts on Ethereum/Quorum designed to represent a derivative and contain all necessary functions required by the lifecycle in the way described in the previous section.
	
	Tokens are represented by the token contract and are issued, owned and provided by the central bank node. Token contracts contain all token functions and keep track of balances of the participants using the tokens.
	
	Finally, the valuation oracle is implemented via another contract. It contains functions to receive requests and getters and setters for valuations provided by the valuation service. The valuation service issues, owns and provides the valuation oracle.

	\subsection{Contract functionality}
	This section lists and describes required functions of each contract used by the Ethereum/Quorum network.

	\subsubsection{Smart derivative contract functionality}

	\begin{lstlisting}
		contract Derivative {
			enum DerivativeStatus { Alive, Initialized, Terminated, Ended, Error }

			mapping (address => uint256) public marginBalance;
			mapping (address => uint256) public premiumBalance;

			modifier onlyParties {}
			modifier onlyAlive {}
			modifier onlyInitialized {}

			function depositP(uint value) public onlyAlive onlyParties returns (bool success) {}
			function depositM(uint value) public onlyAlive onlyInitialized onlyParties returns (bool success) {}
			function debitP(uint value) public onlyAlive onlyInitialized onlyParties returns (bool success) {}
		}
	\end{lstlisting}

	The contract contains several modifiers and mappings used to allow access to functions and to keep track of balances. Functions for the purpose of depositing contract margins and premiums implemented. A settle function allows a party to trigger settlements.

	\subsubsection{Token contract functionality}

	\begin{lstlisting}
		contract Token {
			mapping (adress => uint256) public balanceOf;
			mapping (address => mapping (address => uint256)) public allowance;

			event Transfer(address indexed from, address indexed to, uint256 value);
			event Approval(address indexed _owner, address indexed _spender, uint256 _value);
			event Burn(address indexed from, uint256 value);

			function transfer(address _to, uint256 _value) public returns (bool success) {}
			function transferFrom(address _from, address _to, uint256 value) public returns (bool success) {}

			function approve(address _spender, uint256 _value) public returns (bool success) {}
			function approveAndCall(address _spender, uint256 _value, bytes _extraData) public returns (bool success) {}

			function burn(uint256 _value) public returns (bool success) {}
			function burnFrom(address _from, uint256 _value) public returns (bool success) {}
		}
	\end{lstlisting}

	The contract contains balance and allowance mappings, which are used to keep track of values added to the contract as well as allowances to access a party’s balance. Events such as Transfer, Approval and Burn are used to notify listening parties of a given event. Furthermore, functions in order to transfer, approve and burn token balances are implemented.

	\subsection{Runtime view}
\begin{figure}[hbtp]
	\begin{center}
		\includegraphics[scale=0.35]{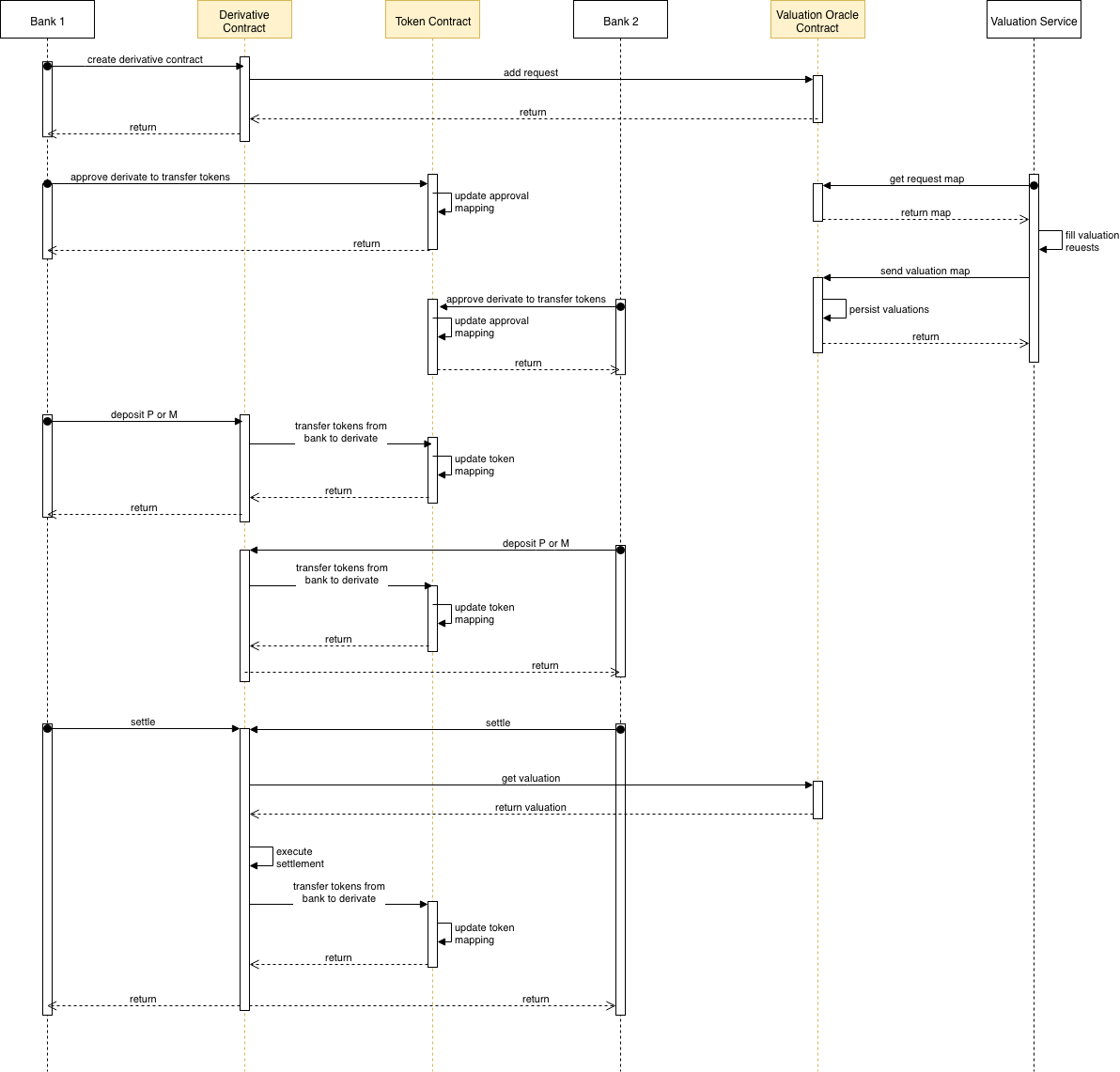}
		\caption[
		]{
			Sequence diagram for the Ethereum smart derivative contract.
		}
		\label{fig:ImplementingSDC:SDCSequenceDiagram}
	\end{center}
	\addtocounter{cpfNumberOfFigures}{1}
	\end{figure}

	Figure~\ref{fig:ImplementingSDC:SDCSequenceDiagram} shows an abstract sequence diagram that visualizes the runtime view of the system.
	
	Each node (white) represents a participant within the network, each contract (yellow) represents a contract object. Ethereum transactions and inter-node-communication (technical, within the system) are not displayed. Notice that either Bank 1 or Bank 2 call the settlement function, not both. Since each single operation, such as the deposition of \(P\) or \(M\) or a settlement is invoked rather than scheduled, the contract needs to be able to determine if a participant is allowed to call a function or in general if requirements are met before executing a function.

\begin{figure}[hbtp]
	\begin{center}
		\includegraphics[scale=0.50]{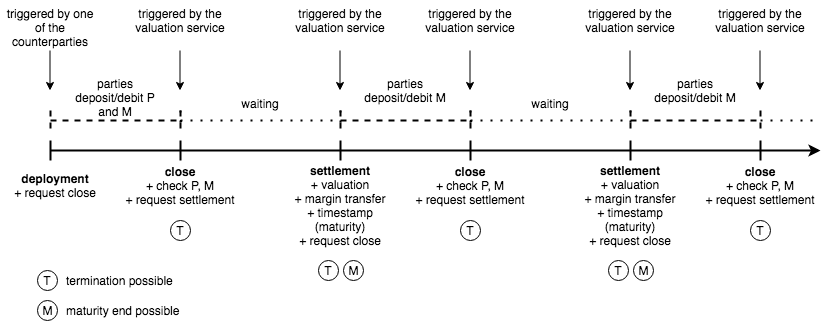}
		\caption[
		]{
			Alternative event time line.
		}
		\label{fig:ImplementingSDC:SDCEthereumTimeLine}
	\end{center}
	\addtocounter{cpfNumberOfFigures}{1}
	\end{figure}

	Figure~\ref{fig:ImplementingSDC:SDCEthereumTimeLine} shows an alternative that relies on the valuation service to act as a timed service in order to trigger settlement of derivatives and deposition of margin buffers.

	\subsection{Critical hypotheses}
	\subsubsection{Stable coin}
	In order to execute the cash flows triggered by the smart derivative contract in Ethereum/Quorum, a cash-on-ledger like coin is needed. This coin ideally needs to be stabilized. In order to do so a central bank or a committee of central banks needs to be implemented that for example backs any coin with real world currency and functions as an exchange to service the conversion of real-world versus digital currency

	\subsubsection{Central services}
	As can be seen in the prototype implementation, a successful and functioning implementation of the smart derivative contract and the corresponding network critically depends on the fact that several well-established and trustworthy central services. Most prominent among them is the valuation service, but also the market data, that is required by the Valuation Oracle and even a time service that equips the naturally timeless Ethereum/Quorum network with time so that execution of settlements and closing of accounts can be scheduled as intended are such third-party services.

	\subsubsection{Private transactions}
	The basic principle of Ethereum requires public transactions in order to enable trustless interaction. To implement private transactions, separate, non-private areas (so-called private
StateDBs in Quorum) have to be implemented within the block chain network. However, the system requires transparency in order to validate any transaction. If a participant is not able to see a transaction, validation and consensus cannot be applied.

	\subsection{Check of requirements}
	Concerning the infrastructure of Ethereum/Quorum, the following previously postulated requirements for the Smart Derivative Contracts are met:

	Margin wallets can be fully implemented using Ethereum/Quorum. Each contract has a unique identifier (address) which allows it to be added to balance mappings or mappings in general. These mappings then function as a kind of ledger.

	Cash-on-ledger/Stable coin is yet an open issue. For the use case of a smart derivative contract representing a real transaction between financial counterparties, a stable coin is needed. The stable coin has to be stabilized either by off-chain collateralization in real currency or similar measures. This requires a central party functioning as the issuer of and exchange for the stable coin.

	Central valuation instance is difficult to achieve since any off-chain information needs to be brought on-chain. In Ethereum/Quorum this can be implemented using Oracle Contracts. An oracle is an on-chain service that is accessed by off-chain services that ultimately provide the required information. Since information brought on-chain as a “fact” by off-chain services cannot be verified by any participant, these services need to be trusted, which contradicts Ethereum’s/Quorum’s trustless design.

	Immutable market data reference is difficult to achieve since any off-chain information needs to be brought on-chain. A market data service can be set up as another oracle Contract that feeds the required market data from a trusted source on the chain. Subsequently, the valuation oracle that values the smart derivative contract can be called using the market data that has been written on the chain as input parameters.

	Timing and Scheduling is difficult to achieve since time is not per se an available entity on block chain networks. Yet another central service has to provide time and might even be responsible to trigger all contract events since smart contracts in Ethereum/Quorum cannot activate or trigger themselves. Any transaction needs to be triggered by a participant or contract. Counterparties however cannot be trusted with timely execution of settlements. Since time is not available within Ethereum/Quorum networks per se, a third party needs to be considered to schedule settlements and closing of accounts.

	Premature termination critically depends on the before mentioned timing and scheduling service to ensure that all defined contract events can be scheduled and triggered on a common time grid. Only then, the smart derivative contract can be executed in the correct chronological order and each participant knows exactly when a settlement or closing of accounts occurs so that the contract can determine whether a termination event has actually occurred at a predefined point on the time grid or not.

	Finality of every contract state is easy to achieve by a contract and poses no issue. However, any contract is stored forever, since in order to keep consistency on-chain, past states and contracts cannot be deleted.

	Privacy is difficult or impossible to achieve by design. Ethereum utilizes trustless transactions by full transparency for any participant. Therefore, basic functions that ensure correct operation on-chain cannot be made secret to a specific group of participants. Quorum has a private ledger. However, one cannot access private information from public states and vice versa. Therefore, Quorum cannot utilize a custom public token for payments while privatizing the underlying contracts.

	\clearpage
	\section{Organizational challenges}
	The most critical aspect of integrating the proposed smart contract architecture is less in direct IT implementation (networking, server infrastructure), but rather in solving the organizational challenges which stem from development and maintenance a decentralized architecture (decentralized development or maintainer model, decentralized application maintenance), as well as the integration of the new decentralized architecture in existing, often proliferated and intricate and of course highly regulated infrastructure of financial institutions. We will now discuss shortly how this task could still be achieved.

Setting up above mentioned infrastructure, consisting of a network of nodes that are run by the individual participants, might be rather inefficient if each party tries to run its node on their respective traditional hardware on premises. This is because each institution and IT provider has its own standards and processes which would then have to be aligned with each other. The more participants (i.e. more nodes) the network has, the more difficult this is likely to become.

A common platform, that allows adding new nodes in a flexible way, together with a blueprint for interested parties, that describes how to technically onboard this platform from their local IT environments, could be helpful in this regard, i.e. lower the IT obstacles for joining the network and therefore help to increase the acceptance of smart derivative contracts in the market.

During a development and testing phase, there seems to be need for a platform that can provide development tools and test infrastructure on demand and that is easily accessible from the local IT infrastructures of the relevant parties (e.g. via VPN).

A promising approach to fulfil these requirements (especially the ones related to development and testing) is making use of cloud-technology. This was already partly done during the prototyping phase by setting up communicating nodes on Amazon Web Services (AWS). Using cloud technology in a real-life environment in banking industries however comes with its own difficulties. While some institutions have been reserved about cloud adoption in the past, many of them seem to be more open by now. Potential benefits regarding flexibility and operating costs, when compared to traditional on-premise IT infrastructure, and progress in the discussions between cloud providers (such as Microsoft Azure, AWS and others), regulators and banks have led to cloud-related projects in many financial institutions \autocite{SherifCloud}.

The details of such a platform and the degree to which cloud-based features could be used for development and testing and later on for a productive environment still have to be investigated. Same applies for the envisioned onboarding blueprint. Open questions range from organizational (e.g. responsibilities for maintaining and operating the platform- infrastructure) to measures to ensure privacy and security of confidential data that might be stored in the network.

Given that cloud activities in many financial institutions are likely to gain more and more momentum, working out the details and actually start developing and testing a cloud-based platform for trading SDCs, looks like an interesting use case that comes at the right time.

	\clearpage
	\section{Integration into existing IT environments}
	We describe all of the following aspects of integration and cross-cutting concepts only regarding the decentralized solution, since a centralized solution can be built with regular architecture and established solutions (e.g. regular web technologies).

	\subsection{Architecture and design patterns}
	\subsubsection{Network topology}
For a decentralized approach, network topology is based around the Ethereum nodes. Every participant deploys its own node in their respective network and whitelists all other nodes for a direct communication.
Oracles are whitelisted for a direct interaction with a single node to publish facts. Interaction between all nodes and a single oracle is not necessary, therefore simplifying the topology.

	\subsubsection{Communication and integration}
All communication is done via the Ethereum protocol, so no separate communication channels need to be established. If additional protection on network layer is required, the communication can be routed through a network of VPN tunnels.

	\subsection{Security and safety}

	\subsubsection{Safety}
To ensure operational safety (failure tolerance) contracts can be verified either through rigorous unit and integration testing or – if possible – through formal verification of correctness.

	\subsubsection{Security}
To ensure security of the smart derivative contract (protection against fraud), the architecture is designed as trustless as possible. Functionality is decentralized, e.g. through escrow functionality of smart contracts. All centralized aspects of the contract remain subject to an attack.
As with regular applications, security of all centralized and decentralized blocks of the application can be increased with a penetration test.

	\subsection{Development concepts}
	\subsubsection{Decentralized development}
Development of the smart derivative contract can be decentralized as practiced in open source projects through established development tools.

For an initial phase of development, a consortium of participants can ensure initial funding and development of a minimum viable product. In this initial phase, a single participant could centralize the development and functional requirement process. Other partners would license the product and indirectly finance the further development.

We subsequently refer to this central development participant as maintainer (analogous to the lead developer of open source projects).

	\subsubsection{Decentralized build, test, deployment process}
In the decentralized case, all participants need to agree on a versioning and deployment process of the decentralized application, as well as the required oracles.

In an initial development phase, the maintainer could establish the build, test and deployment process and prepare for a later decentralized development.

	\subsection{Operational concepts}
	\subsubsection{Application management, permissioning}
Every participant in the network has to establish his own (internal) application management for the node.
On network level, in case of a permissioned network the maintainer could also act as a central (authorizing) administrator until this functionality can be decentralized.

	\subsubsection{Disaster recovery, backup, persistency}
Since the decentralized approach is based on a distributed ledger, persistency is basically provided by the underlying Ethereum network.
For two reasons it may be viable to store data off-chain:
	\begin{itemize}
		\item Storage on a block chain is replicated through the entire network and therefore expensive.
		\item Some data may be private (contract details) or may be required to be off-chain (e.g. additional calculation parameters for the valuation service).
	\end{itemize}
For off-chain data, established storage mechanisms like relational databases are preferred.

	\subsubsection{High availability}
	Redundancy for access to the network can be achieved by connecting multiple nodes to the network. Architecture elements like a central valuation service have to be secured by established high availability mechanisms (redundancy, failover).

	\clearpage
	\section{Conclusion}
	In this work we described how to implement a digital version of a financial derivative contract. The aim within our theoretical concept is to provide a full deterministic definition of a financial derivatives life cycle. The concept provided the basis to implement two prototypes executed on a centralized and distributed ledger platform respectively. We provided a general overview of two block chain infrastructure whereby Ethereum/Quorum has shown to be a solution which can cope with most of our requirements. We gave a detailed technical perspective in terms of infrastructure and implemented functionality.

	In addition, we provided some insights on how such a digital computer protocol can be integrated into existing infrastructures of a financial institution, covering aspects from a procedural, legal and system perspective.

	Several further work and research need to be done to get a smart derivative contract integrated into a real-life context and let that kind of contract become a legal binding character. Especially solving the stable coins question -- to become technically reliable as well as legally -- remains the most essential part.

	However, as several financial institutions and ISDA are currently working on a very precise level on the realization of digital financial derivatives – see e.g. \autocite{IsdaLegal} we are encouraged to push the implementation and integration of our concept into the next phase.
	
	\clearpage
	\printbibliography
\end{document}